\documentclass[]{scrartcl}

\makeatletter
\DeclareOldFontCommand{\rm}{\normalfont\rmfamily}{\mathrm}
\DeclareOldFontCommand{\sf}{\normalfont\sffamily}{\mathsf}
\DeclareOldFontCommand{\tt}{\normalfont\ttfamily}{\mathtt}
\DeclareOldFontCommand{\bf}{\normalfont\bfseries}{\mathbf}
\DeclareOldFontCommand{\it}{\normalfont\itshape}{\mathit}
\DeclareOldFontCommand{\sl}{\normalfont\slshape}{\@nomath\sl}
\DeclareOldFontCommand{\sc}{\normalfont\scshape}{\@nomath\sc}
\makeatother

\usepackage{hyperref}
\usepackage[parfill]{parskip}
\usepackage{siunitx}
\usepackage{cleveref}
\usepackage{todonotes}
\usepackage{scalefnt}
\usepackage{slashed}
\usepackage{microtype}
\usepackage{multirow}
\usepackage{csquotes}
\usepackage{verbatim}
\usepackage{amsmath}
\usepackage{amssymb}

\usepackage{cite}

\usepackage{caption}
\usepackage{subcaption}
\usepackage{graphicx}
\usepackage{nicefrac}

\usepackage{fancyhdr}
\usepackage{authblk}

\newcommand{\abbrev}{\scalefont{.9}}

\newcommand{\NNLO}{\text{\abbrev NNLO}}
\newcommand{\NLO}{\text{\abbrev NLO}}
\newcommand{\LO}{\text{\abbrev LO}}
\newcommand{\EFT}{\text{\abbrev EFT}}

\newcommand{\SM}{\text{\abbrev SM}}
\newcommand{\BSM}{\text{\abbrev BSM}}
\newcommand{\IR}{\text{\abbrev IR}}
\newcommand{\UV}{\text{\abbrev UV}}

\newcommand{\PDF}{\text{\abbrev PDF}}

\newcommand{\LHC}{\text{\abbrev LHC}}
\newcommand{\CTFOURTEEN}{\text{\abbrev CT14}}

\newcommand{\MCFM}{\text{\abbrev MCFM}}

\title{NLO Higgs+jet at large transverse momenta including top quark mass effects}
\author{Tobias Neumann}
\affil{Department of Physics, Illinois Institute of Technology, Chicago, Illinois 60616, USA}
\affil{Fermilab, PO Box 500, Batavia, Illinois 60510, USA}

\fancypagestyle{firstpage}{%
	\lhead{}
	\rhead{IIT-CAPP-18-01}
}

\begin{document}

\maketitle
\thispagestyle{firstpage}

\begin{abstract}
We present a next-to-leading order (\NLO{}) calculation of $H+\text{jet}$ in gluon fusion including the effect of a 
finite top quark 
mass $m_t$ at large transverse momenta. Using the recently published two-loop amplitudes in the high energy expansion 
and our previous setup that 
includes finite $m_t$ effects in a low energy expansion, we are able to obtain $m_t$-finite results for transverse 
momenta below $\SI{225}{\GeV}$ and above $\SI{500}{\GeV}$ with negligible remaining top quark mass uncertainty. The 
only 
remaining region that has to rely on the common leading order (\LO{}) rescaling approach is the threshold region 
$\sqrt{\hat s}\simeq 2m_t$. We demonstrate that this rescaling provides an excellent approximation in the high 
$p_T$ region. Our calculation settles the 
issue of top quark mass effects at large transverse momenta. It is implemented in the parton level Monte Carlo code 
\MCFM{} and is publicly available immediately in version 8.2.
\end{abstract}

\section{Introduction}

With the Higgs boson discovery at the Large Hadron Collider (\LHC{}) \cite{Khachatryan:2014jba,Aad:2015gba}  setting a 
milestone for physics research, the hunt for signals 
beyond those described by the Standard Model (\SM{}) has been more active than ever. Early Higgs studies during Run I, 
limited by statistics and energy, probed rather inclusive properties, and no significant deviations from the \SM{} have 
been found 
\cite{Aad:2015lha,Aad:2015gba,Aad:2015mxa,Khachatryan:2014jba,Aad:2015zhl,Khachatryan:2016vau,Giardino:2013bma}. 
Differential Higgs measurements 
\cite{Aad:2016lvc,Aad:2015tna,Aad:2014tca,Aad:2014lwa,Khachatryan:2015yvw,Khachatryan:2015rxa,Khachatryan:2016vnn} 
testing the \SM{} were limited by statistics rather than theory predictions.
Recent experimental analyses consider the Higgs boson in a highly boosted regime with transverse momenta ($p_T$) of 
$\SI{450}{\GeV}$ to $\SI{1}{\TeV}$ \cite{CMS:2017cbv,Sirunyan:2017dgc,Vernieri:2017jqy}. Clearly there is an evolving 
need for most precise predictions in differential quantities and in the high energy regime that needs to be filled.

To constrain physics beyond the Standard Model (\BSM{}) using gluon fusion Higgs production, one of the most promising 
approaches is to consider large transverse energies 
\cite{Harlander:2013oja,Dawson:2014ora,Dawson:2015gka,Banfi:2013yoa,Azatov:2013xha,Grojean:2013nya,
	Schlaffer:2014osa,Buschmann:2014twa,Buschmann:2014sia,Langenegger:2015lra,
	Ghosh:2014wxa,Grazzini:2016paz,Cohen:2017rsk}.
Unfortunately an effective field theory (\EFT{}) description that integrates out the top quark as a heavy particle has 
to 
be used at 
higher perturbative orders to approximate the complicated massive loop integrals. The operator used in the 
infinite top quark mass \EFT{} is the same operator that constitutes the leading \EFT{} operator to search for new 
physics. 
Thus 
the only 
reliable way to directly disentangle 
\SM{} gluon fusion from heavy \BSM{} loop contributions requires computing the full top quark mass dependence at large 
energies. 
It is this region, in which finite top quark mass effects are unconstrained at \NLO{}, that motivates our 
study.\footnote{While finalizing our manuscript we became aware of a study similar to 
	ours in 
	ref.~\cite{Lindert:2018iug}. Additionally, 
	results based on a fully numerical evaluation of the two-loop integrals with full top quark mass dependence have 
	been 
	presented in ref.~\cite{Jones:2018hbb}. We 
	still believe that our study and implementation are useful for large transverse momenta, where the difference to 
	the 
	full calculation should be negligible and the full calculation seems to be numerically challenging for large 
	transverse 
	momenta.} When 
it comes 
to 
precision \SM{} predictions to constrain \BSM{} physics, not just the size of the perturbative corrections to Higgs 
production are  
of importance, which are well described by the \EFT{}, but also the shape of transverse momentum 
distributions, for example. In this study we can also evaluate whether top quark mass effects distort the shape of the 
Higgs 
high 
transverse momentum distribution compared to previous approximations.

While gluon fusion induced Higgs production mediated through a massive top quark loop was calculated at \LO{} a long 
time ago \cite{Ellis:1987xu}\footnote{For a recent overview of Higgs production and decay cross sections we refer to 
ref.~\cite{Spira:2016ztx}.}, the 
difficulty is considerably amplified at \NLO{} and in Higgs production with a jet, where massive two-loop amplitudes 
have to be calculated. An efficient analytical 
evaluation of these integrals for Higgs+jet is currently not within reach 
\cite{Bonciani:2016qxi,Boggia:2017hyq}. Fortunately the \EFT{} approach turns out to provide an excellent 
approximation of perturbative corrections even 
for differential Higgs+jet quantities \cite{Harlander:2012hf,Neumann:2014nha}. This was shown using a low energy 
expansion below the top quark threshold $p_T\ll 2m_T$, and as such can not constrain the effects of a finite top quark 
mass for large energies. Going beyond estimations of top quark mass effects, studies with direct predictions including 
them 
were performed \cite{Banfi:2013eda,Hamilton:2015nsa,Frederix:2016cnl,Neumann:2016dny}, but were always limited by low 
energy 
approximations. And only the \EFT{} approach, which reduces the needed number of loops to evaluate by one, allowed for 
an 
evaluation of 
Higgs+jet at \NNLO{} \cite{Boughezal:2013uia,Chen:2014gva,Boughezal:2015aha,Boughezal:2015dra,Chen:2016zka}. Residual 
perturbative uncertainties estimated by renormalization and factorization scale variation are estimated to be about 
$10\%$ at \NNLO{}, cutting in half the estimated \NLO{} uncertainty.

Studies targeting top quark mass effects specifically in the large transverse momentum region of Higgs production were 
performed using resummation \cite{Caola:2016upw}, and by using a factorization of the mass scale from $p_T$ in the high 
$p_T$ limit \cite{Braaten:2017lxx}. In the latter case the developed formalism was only applied to the subprocess 
$q\bar{q}$ at leading order in $1/p_T$ and leading order in $\alpha_s$. The former study only improves the \NLO{} 
calculation 
through a rescaling $k$-factor and not directly, as their high energy approximation was shown to deviate from the exact 
result at 
\LO{}. Their suggested approach is to rescale exact \LO{} results by a $k$-factor obtained in their high energy 
approximation. A different approach is to match different hard-jet multiplicities and parton showers 
\cite{Buschmann:2014sia,Frederix:2016cnl}.

It is the goal of this study to extend our previous setup \cite{Neumann:2016dny}, publicly available in \MCFM{} 
\cite{Campbell:1999ah,Campbell:2011bn,Campbell:2015qma}, and which provides finite 
$m_t$ effects 
in the region $p_t\ll 2m_t$, to predictions with a finite $m_t$ for $p_T\gg 2 m_t$. We also study the validity of 
the 
common \LO{} rescaling approach in the region of high $p_T$. To do this we implement the recently published two-loop 
amplitudes in the high energy expansion \cite{Kudashkin:2017skd}.

\section{Calculation}
\label{sec:calculation}
Our calculation is based on the existing \NLO{} Higgs+jet setup in \MCFM{}-8.1 
\cite{Neumann:2016dny}. It uses an 
asymptotic expansion in $\Lambda/(2m_t)$ only for the finite part of the virtual two-loop 
amplitudes, but is exact in the top quark mass otherwise. Here $\Lambda$ is a placeholder for all 
kinematical scales of the process. For Higgs $p_T$ smaller than $\simeq \SI{225}{\GeV}$ the asymptotic expansion was 
shown to be convergent and provides an excellent approximation of the full top quark mass dependence. For 
energies larger than $\simeq\SI{300}{\GeV}$ the expansion breaks down and finite top quark mass effects could become 
larger 
than 
$8\%$, such that either a full calculation is necessary or another approximation is needed for sufficiently large 
$p_T$. Here, we 
fill this gap for the latter case.

We have implemented the one- and two-loop Higgs plus three parton helicity amplitudes in the high energy expansion from 
ref.~\cite{Kudashkin:2017skd}. The expansion is performed in $\kappa\equiv -\left(m_t^2/\hat{s}\right)^k$ to order 
$k=1$ while 
retaining an expansion in $\eta\equiv -\left(m_H^2/(4m_t^2)\right)^l$ only to first order $l=0$. Here $m_H$ is the 
Higgs mass and $\hat{s}$ the partonic center of mass energy. The amplitudes are 
given as the finite parts after \UV{} renormalization and Catani \IR{} subtraction \cite{Catani:1998bh} using 
$d=4-2\epsilon$ Born one-loop amplitudes. We have performed a conversion to the 't\,Hooft-Veltman scheme for use in 
\MCFM{} and additionally restored the renormalization scale dependence.

At \LO{} Higgs+jet relies on one-loop amplitudes and is known with the exact top quark mass dependence, which allows us 
to 
compare it with the result from the high energy expansion. This gives an estimate on how far we can trust 
the approximation when using the two-loop amplitudes. Having established trust in the validity of the two-loop 
amplitudes in the high $p_T$ region we can then compare the results with the Born-rescaling approximation. In this 
rescaling approximation the 
finite part of the two-loop virtual amplitude is point-wise rescaled by the Born amplitude in the full theory divided 
by the Born amplitude in the \EFT{}.

For our study we choose a center of mass energy of $\sqrt{s}=\SI{13}{\TeV}$ and a common renormalization and 
factorization 
scale of $\mu_R=\mu_F=\sqrt{m_H^2+p_T^2}$, where $m_H=\SI{125.0}{\GeV}$ and $p_T$ is the Higgs 
transverse momentum. Although the region of high $p_T$ motivates using the six-flavor scheme, no matching parton 
distribution functions (\PDF{}s) are available. So for consistency we work in the five-flavor scheme with an 
on-shell 
top quark mass of $m_\text{t}=\SI{173.2}{\GeV}$. We use \CTFOURTEEN{} \PDF{}s \cite{Dulat:2015mca} at \NLO{} accuracy 
for the \NLO{} cross 
section and at \LO{} accuracy for our \LO{} results. The value of $\alpha_s$ is given at the according order by the 
\PDF{} set. Finally, we use the anti-$k_T$ jet algorithm with $p_{T,\text{jet}}>\SI{30}{\GeV}$, 
$|\eta_\text{jet}|<2.4$ and $R=0.5$.

\section{Results}

\begin{figure}
	\includegraphics[width=\columnwidth]{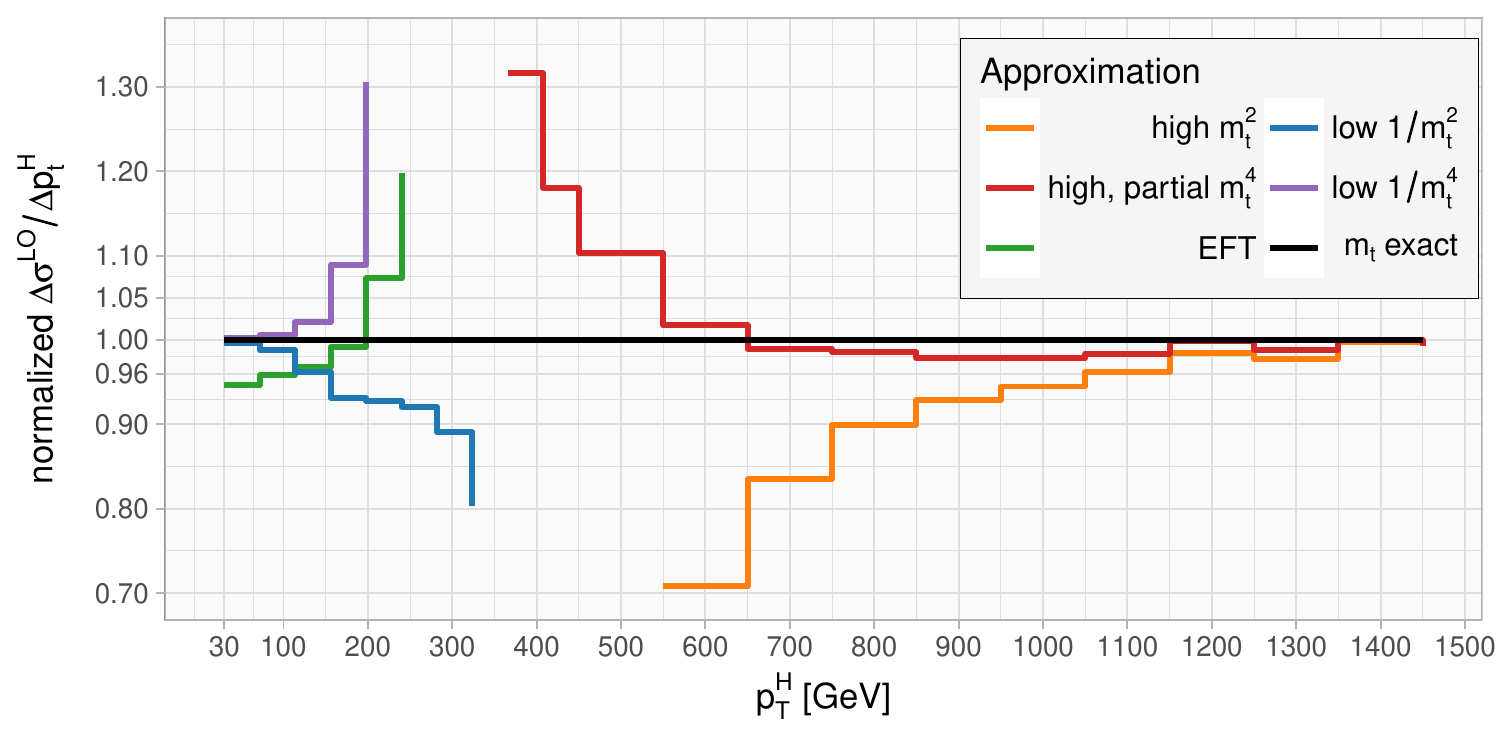} 
	\caption{Normalized Higgs transverse momentum distribution at \LO{} in the low energy $1/m_t$ expansion 
	for low 
	$p_T$, in the large energy expansion for large $p_T$ and with exact $m_t$ dependence as normalization. The first 
	order of the large energy expansion deviates by more than $30\%$ and is not shown.}
	\label{fig:locomparison}
\end{figure}

The first question one has to ask is in how far one can trust the two-loop high energy amplitudes to describe the exact 
$m_t$ dependence. In lack of the $m_t$-exact two-loop amplitudes for comparison, one has to resort to a different 
method to evaluate this trust. For example, one could observe a convergent behavior of the expansion, but 
this would require some higher expansion order than is available as we will see below. Instead we can study how well 
the 
expansion works at \LO{}. This has some limitations though, as will be discussed below. To study the high energy 
expansion we consider \cref{fig:locomparison}: Shown is the \LO{} Higgs
transverse momentum distribution in various approximations normalized to the distribution with exact $m_t$-dependence. 
The approximations shown are the low energy expansions up to order $1/m_t^k$ for $k=0,2,4$, where $k=0$ describes the 
\EFT{}, as well as the high energy expansion.

\begin{figure}
	\includegraphics[width=\columnwidth]{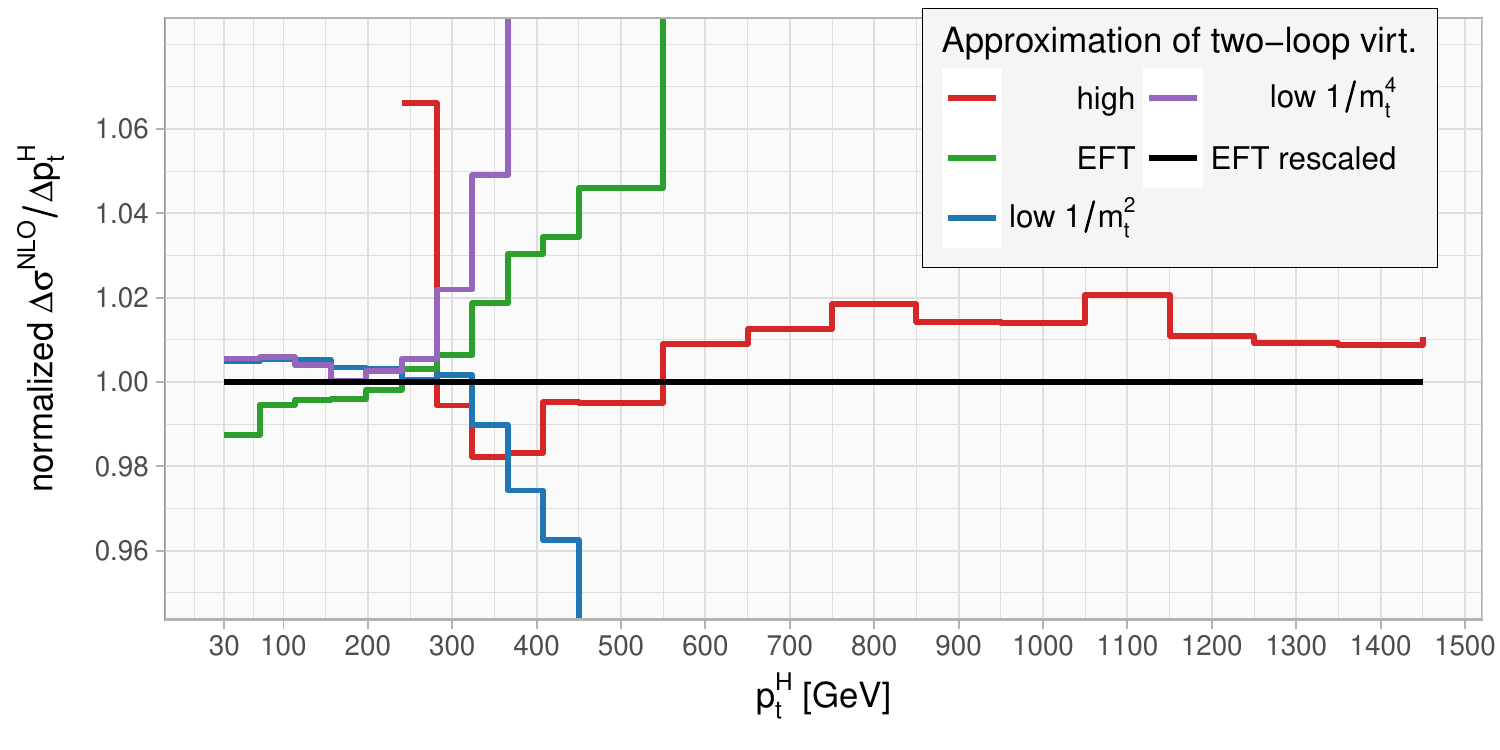} 
	\caption{Normalized Higgs transverse momentum distribution at \NLO{}. Born and real emission contributions are 
	taken into account $m_t$-exactly and only the finite part of the two-loop virtual correction is approximated in 
	various 
	ways: In the low $p_T$ region the low energy $1/m_t$ expansion is applied while in the high $p_T$ region the high 
	energy approximation is used. The rescaled result takes the finite part of the two-loop \EFT{} amplitude 
	(coinciding 
	with $1/m_t^0$) and rescales it by the ratio of the Born amplitude in full theory to the \EFT{} amplitude.  }
	\label{fig:nlo}
\end{figure}

For the low energy expansion the convergence is poor and is practically non-existent beyond $\simeq\SI{100}{\GeV}$. At 
\NLO{} 
though, using only the expansion in the two-loop 
amplitude, the region of convergence increases to about \SI{250}{\GeV} as shown in 
ref.~\cite{Neumann:2016dny}. This can also be seen in \cref{fig:nlo}. A simple explanation is that for 
two-loop diagrams the topology does not force all center 
of mass energy to go through the top quark loop, such that $\sim 2 m_t$ threshold effects are further washed out.

The amplitudes in the high energy expansion are given up to order $\kappa^1\equiv(m_t^2/\hat{s})^1$. Naively using them 
for the \LO{} cross section includes partial effects of order $\kappa^2$. This is labeled in the plot as \enquote{high, 
partial $m_t^4$}. Only including $m_t^2$ terms in the 
cross section is labeled with \enquote{high $m_t^2$}. By the same argument given above, one would expect the high 
energy 
approximation to work better for one-loop diagrams than for two-loop diagrams. Nevertheless the difference between
using the full $\mathcal{O}(\kappa^1)$ amplitudes and the $m_t$-exact result is less than two percent beyond 
\SI{500}{\GeV}, which 
gives motivation to trust that the two-loop high energy amplitudes describe the full top quark mass dependence at a 
similar 
level.  Considering that the \NLO{} scale uncertainty is 
about $20\%$ \cite{Neumann:2016dny} and still about $10\%$ at 
\NNLO{} \cite{Boughezal:2015aha,Chen:2016zka}, any remaining top quark mass uncertainty can then be considered 
negligible. Ideally a more precise estimate could be established by 
including full $\mathcal{O}(\kappa^2)$ and $\mathcal{O}(\kappa^3)$ terms for the one- and two-loop amplitudes.

Having shown that the large energy expansion describes the full \LO{} result at percent level accuracy beyond 
$\SI{500}{\GeV}$, we expect a similar behavior for the two-loop 
amplitudes. At \NLO{} the two-loop amplitudes additionally only enter as the virtual corrections and a bulk of the 
perturbative corrections at large $p_T$ comes from the real emission which we include with full $m_t$ dependence. The 
error from using the large energy expansion estimated at \LO{} should thus be conservative.

At \NLO{} we show the Higgs $p_T$ distribution in \cref{fig:nlo}, where, to emphasize again, only the finite part of 
the two-loop virtual corrections is not exact in $m_t$ and is approximated in different ways. It is obtained using 
either a 
$1/m_t$ expansion in the region of small 
$p_T$, or in the high energy expansion up to order $\kappa^1$ for large $p_T$. Additionally, using the rescaling 
approach as described in 
\cref{sec:calculation} we
obtain an approximation that can be used over the whole range of $p_T$. The latter approach was used for example in 
ref.~\cite{Frederix:2016cnl} and shown to agree with the low energy asymptotic expansion at the 
percent level for $p_T\lesssim \SI{225}{\GeV}$ \cite{Neumann:2016dny}.

Since we are strongly interested in possible shape corrections due to using a finite top quark mass, we present the 
distribution normalized to the rescaling approximation result. This shows the 
additional corrections compared to the previous best approximation at high $p_T$ in percent. The scale variation 
uncertainty of about \SI{20}{\%} changes only little with respect to the \EFT{} result and other approximations and can 
be found for example in our previous study \cite{Neumann:2016dny}.

The low energy expansion indeed extends its convergent behavior to about \SI{250}{\GeV} with corrections of less than a 
percent compared to the rescaling approach. The high energy expansion, 
which we believe approximates the full result by better than 
$2\%$, is consistent with the corrections at low $p_T$ and  increases the cross section by about $1-2\%$ compared to the
rescaled result. It is remarkable that these top quark mass corrections are flat 
within $1\%$ over the whole range of large $p_T$. In this sense one is free to 
choose either the rescaling approach or the high energy approximation. Nevertheless the high energy approximation is a 
systematic approach, whereas the rescaling approach was done ad hoc without prior validation. We thus recommend to use 
the high energy approximation for transverse momenta beyond $\sim\SI{500}{\GeV}$.

\section{Conclusions}

We have presented a \NLO{} calculation of Higgs+jet with negligible remaining top quark mass 
uncertainty in the region of low transverse momenta 
$p_T\lesssim \SI{225}{\GeV}$ \cite{Neumann:2016dny} and, as shown in this analysis, also in the region of large 
transverse momenta $p_T\gtrsim\SI{500}{\GeV}$. We have demonstrated that the approach of rescaling the finite part of 
the 
two-loop virtual amplitude by the $m_t$-exact Born amplitude approximates the $m_t$-exact \NLO{} result by better than 
a few 
percent.
The elimination of the top-quark mass uncertainty at high $p_T$ at \NLO{} now allows one to rescale \NNLO{} results 
obtained in the \EFT{} by a \NLO{} $k$-factor $\NLO{}(m_t) / \NLO{}(\EFT{})$. Our implementation is publicly available 
immediately in \MCFM{}-8.2.

\paragraph{Acknowledgments.}

We would like to thank Christopher Wever for providing us with the terms of order 
$\epsilon^2$ and a phase space point evaluation of the high energy 
one-loop amplitude of ref.~\cite{Kudashkin:2017skd}. We would like to thank John 
Campbell for extensive discussion and comments on the manuscript as well as Zack Sullivan for valuable discussion and 
comments on the manuscript and Stefan Prestel for helpful discussion.

This work was supported by the U.S.\ Department of Energy under award No.\ DE-SC0008347. This document was prepared 
using the resources of the Fermi National Accelerator Laboratory (Fermilab), a U.S. Department of Energy, Office of 
Science, HEP User Facility. Fermilab is managed by Fermi Research Alliance, LLC (FRA), acting under Contract No.\ 
DE-AC02-07CH11359.

\bibliographystyle{JHEP}
\bibliography{hhpt}

\providecommand{\href}[2]{#2}\begingroup\raggedright\begin{thebibliography}{10}

\bibitem{Khachatryan:2014jba}
{\scshape CMS} collaboration, V.~Khachatryan et~al., \emph{{Precise
  determination of the mass of the Higgs boson and tests of compatibility of
  its couplings with the standard model predictions using proton collisions at
  7 and 8 $\,\text {TeV}$}},
  \href{https://doi.org/10.1140/epjc/s10052-015-3351-7}{\emph{Eur. Phys. J.}
  {\bfseries C75} (2015) 212},
  [\href{https://arxiv.org/abs/1412.8662}{{\ttfamily 1412.8662}}].

\bibitem{Aad:2015gba}
{\scshape ATLAS} collaboration, G.~Aad et~al., \emph{{Measurements of the Higgs
  boson production and decay rates and coupling strengths using pp collision
  data at $\sqrt{s}=7$ and 8 TeV in the ATLAS experiment}},
  \href{https://doi.org/10.1140/epjc/s10052-015-3769-y}{\emph{Eur. Phys. J.}
  {\bfseries C76} (2016) 6},
  [\href{https://arxiv.org/abs/1507.04548}{{\ttfamily 1507.04548}}].

\bibitem{Aad:2015lha}
{\scshape ATLAS} collaboration, G.~Aad et~al., \emph{{Measurements of the Total
  and Differential Higgs Boson Production Cross Sections Combining the H→γγ
  and H→ZZ*→4ℓ Decay Channels at $\sqrt{s}$=8  TeV with the ATLAS
  Detector}}, \href{https://doi.org/10.1103/PhysRevLett.115.091801}{\emph{Phys.
  Rev. Lett.} {\bfseries 115} (2015) 091801},
  [\href{https://arxiv.org/abs/1504.05833}{{\ttfamily 1504.05833}}].

\bibitem{Aad:2015mxa}
{\scshape ATLAS} collaboration, G.~Aad et~al., \emph{{Study of the spin and
  parity of the Higgs boson in diboson decays with the ATLAS detector}},
  \href{https://doi.org/10.1140/epjc/s10052-015-3685-1,
  10.1140/epjc/s10052-016-3934-y}{\emph{Eur. Phys. J.} {\bfseries C75} (2015)
  476}, [\href{https://arxiv.org/abs/1506.05669}{{\ttfamily 1506.05669}}].

\bibitem{Aad:2015zhl}
{\scshape ATLAS, CMS} collaboration, G.~Aad et~al., \emph{{Combined Measurement
  of the Higgs Boson Mass in $pp$ Collisions at $\sqrt{s}=7$ and 8 TeV with the
  ATLAS and CMS Experiments}},
  \href{https://doi.org/10.1103/PhysRevLett.114.191803}{\emph{Phys. Rev. Lett.}
  {\bfseries 114} (2015) 191803},
  [\href{https://arxiv.org/abs/1503.07589}{{\ttfamily 1503.07589}}].

\bibitem{Khachatryan:2016vau}
{\scshape ATLAS, CMS} collaboration, G.~Aad et~al., \emph{{Measurements of the
  Higgs boson production and decay rates and constraints on its couplings from
  a combined ATLAS and CMS analysis of the LHC pp collision data at $
  \sqrt{s}=7 $ and 8 TeV}},
  \href{https://doi.org/10.1007/JHEP08(2016)045}{\emph{JHEP} {\bfseries 08}
  (2016) 045}, [\href{https://arxiv.org/abs/1606.02266}{{\ttfamily
  1606.02266}}].

\bibitem{Giardino:2013bma}
P.~P. Giardino, K.~Kannike, I.~Masina, M.~Raidal and A.~Strumia, \emph{{The
  universal Higgs fit}},
  \href{https://doi.org/10.1007/JHEP05(2014)046}{\emph{JHEP} {\bfseries 05}
  (2014) 046}, [\href{https://arxiv.org/abs/1303.3570}{{\ttfamily 1303.3570}}].

\bibitem{Aad:2016lvc}
{\scshape ATLAS} collaboration, G.~Aad et~al., \emph{{Measurement of fiducial
  differential cross sections of gluon-fusion production of Higgs bosons
  decaying to WW$^{∗}$→eνμν with the ATLAS detector at $ \sqrt{s}=8 $
  TeV}}, \href{https://doi.org/10.1007/JHEP08(2016)104}{\emph{JHEP} {\bfseries
  08} (2016) 104}, [\href{https://arxiv.org/abs/1604.02997}{{\ttfamily
  1604.02997}}].

\bibitem{Aad:2015tna}
{\scshape ATLAS} collaboration, G.~Aad et~al., \emph{{Constraints on
  non-Standard Model Higgs boson interactions in an effective Lagrangian using
  differential cross sections measured in the $H \rightarrow \gamma\gamma$
  decay channel at $\sqrt{s} = 8$TeV with the ATLAS detector}},
  \href{https://doi.org/10.1016/j.physletb.2015.11.071}{\emph{Phys. Lett.}
  {\bfseries B753} (2016) 69--85},
  [\href{https://arxiv.org/abs/1508.02507}{{\ttfamily 1508.02507}}].

\bibitem{Aad:2014tca}
{\scshape ATLAS} collaboration, G.~Aad et~al., \emph{{Fiducial and differential
  cross sections of Higgs boson production measured in the four-lepton decay
  channel in $pp$ collisions at $\sqrt{s}$=8 TeV with the ATLAS detector}},
  \href{https://doi.org/10.1016/j.physletb.2014.09.054}{\emph{Phys. Lett.}
  {\bfseries B738} (2014) 234--253},
  [\href{https://arxiv.org/abs/1408.3226}{{\ttfamily 1408.3226}}].

\bibitem{Aad:2014lwa}
{\scshape ATLAS} collaboration, G.~Aad et~al., \emph{{Measurements of fiducial
  and differential cross sections for Higgs boson production in the diphoton
  decay channel at $\sqrt{s}=8$ TeV with ATLAS}},
  \href{https://doi.org/10.1007/JHEP09(2014)112}{\emph{JHEP} {\bfseries 09}
  (2014) 112}, [\href{https://arxiv.org/abs/1407.4222}{{\ttfamily 1407.4222}}].

\bibitem{Khachatryan:2015yvw}
{\scshape CMS} collaboration, V.~Khachatryan et~al., \emph{{Measurement of
  differential and integrated fiducial cross sections for Higgs boson
  production in the four-lepton decay channel in pp collisions at $ \sqrt{s}=7
  $ and 8 TeV}}, \href{https://doi.org/10.1007/JHEP04(2016)005}{\emph{JHEP}
  {\bfseries 04} (2016) 005},
  [\href{https://arxiv.org/abs/1512.08377}{{\ttfamily 1512.08377}}].

\bibitem{Khachatryan:2015rxa}
{\scshape CMS} collaboration, V.~Khachatryan et~al., \emph{{Measurement of
  differential cross sections for Higgs boson production in the diphoton decay
  channel in pp collisions at $\sqrt{s}=8\,\text {TeV} $}},
  \href{https://doi.org/10.1140/epjc/s10052-015-3853-3}{\emph{Eur. Phys. J.}
  {\bfseries C76} (2016) 13},
  [\href{https://arxiv.org/abs/1508.07819}{{\ttfamily 1508.07819}}].

\bibitem{Khachatryan:2016vnn}
{\scshape CMS} collaboration, V.~Khachatryan et~al., \emph{{Measurement of the
  transverse momentum spectrum of the Higgs boson produced in pp collisions at
  $ \sqrt{s}=8 $ TeV using $H \to WW$ decays}},
  \href{https://doi.org/10.1007/JHEP03(2017)032}{\emph{JHEP} {\bfseries 03}
  (2017) 032}, [\href{https://arxiv.org/abs/1606.01522}{{\ttfamily
  1606.01522}}].

\bibitem{CMS:2017cbv}
{\scshape CMS} collaboration, C.~Collaboration, \emph{{Inclusive search for the
  standard model Higgs boson produced in pp collisions at
  $\sqrt{s}=13~\mathrm{TeV}$ using H$\rightarrow \mathrm{b\bar{\mathrm{b}}}$
  decays}}, .

\bibitem{Sirunyan:2017dgc}
{\scshape CMS} collaboration, A.~M. Sirunyan et~al., \emph{{Inclusive search
  for a highly boosted Higgs boson decaying to a bottom quark-antiquark pair}},
   \href{https://arxiv.org/abs/1709.05543}{{\ttfamily 1709.05543}}.

\bibitem{Vernieri:2017jqy}
{\scshape CMS} collaboration, C.~Vernieri, \emph{{Inclusive Search for Boosted
  Higgs Bosons Using H$ \rightarrow \mathrm{b\overline{b}}$ Decays with the CMS
  Experiment}}, {\emph{PoS} {\bfseries EPS-HEP2017} (2017) 349},
  [\href{https://arxiv.org/abs/1711.10508}{{\ttfamily 1711.10508}}].

\bibitem{Harlander:2013oja}
R.~V. Harlander and T.~Neumann, \emph{{Probing the nature of the Higgs-gluon
  coupling}}, \href{https://doi.org/10.1103/PhysRevD.88.074015}{\emph{Phys.
  Rev.} {\bfseries D88} (2013) 074015},
  [\href{https://arxiv.org/abs/1308.2225}{{\ttfamily 1308.2225}}].

\bibitem{Dawson:2014ora}
S.~Dawson, I.~M. Lewis and M.~Zeng, \emph{{Effective field theory for Higgs
  boson plus jet production}},
  \href{https://doi.org/10.1103/PhysRevD.90.093007}{\emph{Phys. Rev.}
  {\bfseries D90} (2014) 093007},
  [\href{https://arxiv.org/abs/1409.6299}{{\ttfamily 1409.6299}}].

\bibitem{Dawson:2015gka}
S.~Dawson, I.~M. Lewis and M.~Zeng, \emph{{Usefulness of effective field theory
  for boosted Higgs production}},
  \href{https://doi.org/10.1103/PhysRevD.91.074012}{\emph{Phys. Rev.}
  {\bfseries D91} (2015) 074012},
  [\href{https://arxiv.org/abs/1501.04103}{{\ttfamily 1501.04103}}].

\bibitem{Banfi:2013yoa}
A.~Banfi, A.~Martin and V.~Sanz, \emph{{Probing top-partners in Higgs+jets}},
  \href{https://doi.org/10.1007/JHEP08(2014)053}{\emph{JHEP} {\bfseries 08}
  (2014) 053}, [\href{https://arxiv.org/abs/1308.4771}{{\ttfamily 1308.4771}}].

\bibitem{Azatov:2013xha}
A.~Azatov and A.~Paul, \emph{{Probing Higgs couplings with high $p_T$ Higgs
  production}}, \href{https://doi.org/10.1007/JHEP01(2014)014}{\emph{JHEP}
  {\bfseries 01} (2014) 014},
  [\href{https://arxiv.org/abs/1309.5273}{{\ttfamily 1309.5273}}].

\bibitem{Grojean:2013nya}
C.~Grojean, E.~Salvioni, M.~Schlaffer and A.~Weiler, \emph{{Very boosted Higgs
  in gluon fusion}}, \href{https://doi.org/10.1007/JHEP05(2014)022}{\emph{JHEP}
  {\bfseries 05} (2014) 022},
  [\href{https://arxiv.org/abs/1312.3317}{{\ttfamily 1312.3317}}].

\bibitem{Schlaffer:2014osa}
M.~Schlaffer, M.~Spannowsky, M.~Takeuchi, A.~Weiler and C.~Wymant,
  \emph{{Boosted Higgs Shapes}},
  \href{https://doi.org/10.1140/epjc/s10052-014-3120-z}{\emph{Eur. Phys. J.}
  {\bfseries C74} (2014) 3120},
  [\href{https://arxiv.org/abs/1405.4295}{{\ttfamily 1405.4295}}].

\bibitem{Buschmann:2014twa}
M.~Buschmann, C.~Englert, D.~Goncalves, T.~Plehn and M.~Spannowsky,
  \emph{{Resolving the Higgs-Gluon Coupling with Jets}},
  \href{https://doi.org/10.1103/PhysRevD.90.013010}{\emph{Phys. Rev.}
  {\bfseries D90} (2014) 013010},
  [\href{https://arxiv.org/abs/1405.7651}{{\ttfamily 1405.7651}}].

\bibitem{Buschmann:2014sia}
M.~Buschmann, D.~Goncalves, S.~Kuttimalai, M.~Schonherr, F.~Krauss and
  T.~Plehn, \emph{{Mass Effects in the Higgs-Gluon Coupling: Boosted vs
  Off-Shell Production}},
  \href{https://doi.org/10.1007/JHEP02(2015)038}{\emph{JHEP} {\bfseries 02}
  (2015) 038}, [\href{https://arxiv.org/abs/1410.5806}{{\ttfamily 1410.5806}}].

\bibitem{Langenegger:2015lra}
U.~Langenegger, M.~Spira and I.~Strebel, \emph{{Testing the Higgs Boson
  Coupling to Gluons}},  \href{https://arxiv.org/abs/1507.01373}{{\ttfamily
  1507.01373}}.

\bibitem{Ghosh:2014wxa}
D.~Ghosh and M.~Wiebusch, \emph{{Dimension-six triple gluon operator in
  Higgs$+$jet observables}},
  \href{https://doi.org/10.1103/PhysRevD.91.031701}{\emph{Phys. Rev.}
  {\bfseries D91} (2015) 031701},
  [\href{https://arxiv.org/abs/1411.2029}{{\ttfamily 1411.2029}}].

\bibitem{Grazzini:2016paz}
M.~Grazzini, A.~Ilnicka, M.~Spira and M.~Wiesemann, \emph{{Modeling BSM effects
  on the Higgs transverse-momentum spectrum in an EFT approach}},
  \href{https://doi.org/10.1007/JHEP03(2017)115}{\emph{JHEP} {\bfseries 03}
  (2017) 115}, [\href{https://arxiv.org/abs/1612.00283}{{\ttfamily
  1612.00283}}].

\bibitem{Cohen:2017rsk}
J.~Cohen, S.~Bar-Shalom, G.~Eilam and A.~Soni, \emph{{Light-quarks Yukawa and
  new physics in exclusive high-$p_T$ Higgs + jet(b-jet) events}},
  \href{https://arxiv.org/abs/1705.09295}{{\ttfamily 1705.09295}}.

\bibitem{Lindert:2018iug}
J.~M. Lindert, K.~Kudashkin, K.~Melnikov and C.~Wever, \emph{{Higgs bosons with
  large transverse momentum at the LHC}},
  \href{https://arxiv.org/abs/1801.08226}{{\ttfamily 1801.08226}}.

\bibitem{Jones:2018hbb}
S.~P. Jones, M.~Kerner and G.~Luisoni, \emph{{NLO QCD corrections to Higgs
  boson plus jet production with full top-quark mass dependence}},
  \href{https://arxiv.org/abs/1802.00349}{{\ttfamily 1802.00349}}.

\bibitem{Ellis:1987xu}
R.~K. Ellis, I.~Hinchliffe, M.~Soldate and J.~J. van~der Bij, \emph{{Higgs
  Decay to tau+ tau-: A Possible Signature of Intermediate Mass Higgs Bosons at
  the SSC}}, \href{https://doi.org/10.1016/0550-3213(88)90019-3}{\emph{Nucl.
  Phys.} {\bfseries B297} (1988) 221--243}.

\bibitem{Spira:2016ztx}
M.~Spira, \emph{{Higgs Boson Production and Decay at Hadron Colliders}},
  \href{https://doi.org/10.1016/j.ppnp.2017.04.001}{\emph{Prog. Part. Nucl.
  Phys.} {\bfseries 95} (2017) 98--159},
  [\href{https://arxiv.org/abs/1612.07651}{{\ttfamily 1612.07651}}].

\bibitem{Bonciani:2016qxi}
R.~Bonciani, V.~Del~Duca, H.~Frellesvig, J.~M. Henn, F.~Moriello and V.~A.
  Smirnov, \emph{{Two-loop planar master integrals for Higgs$\to 3$ partons
  with full heavy-quark mass dependence}},
  \href{https://doi.org/10.1007/JHEP12(2016)096}{\emph{JHEP} {\bfseries 12}
  (2016) 096}, [\href{https://arxiv.org/abs/1609.06685}{{\ttfamily
  1609.06685}}].

\bibitem{Boggia:2017hyq}
M.~Boggia et~al., \emph{{The HiggsTools Handbook: Concepts and observables for
  deciphering the Nature of the Higgs Sector}},
  \href{https://arxiv.org/abs/1711.09875}{{\ttfamily 1711.09875}}.

\bibitem{Harlander:2012hf}
R.~V. Harlander, T.~Neumann, K.~J. Ozeren and M.~Wiesemann, \emph{{Top-mass
  effects in differential Higgs production through gluon fusion at order
  $\alpha_s^4$}}, \href{https://doi.org/10.1007/JHEP08(2012)139}{\emph{JHEP}
  {\bfseries 08} (2012) 139},
  [\href{https://arxiv.org/abs/1206.0157}{{\ttfamily 1206.0157}}].

\bibitem{Neumann:2014nha}
T.~Neumann and M.~Wiesemann, \emph{{Finite top-mass effects in gluon-induced
  Higgs production with a jet-veto at NNLO}},
  \href{https://doi.org/10.1007/JHEP11(2014)150}{\emph{JHEP} {\bfseries 11}
  (2014) 150}, [\href{https://arxiv.org/abs/1408.6836}{{\ttfamily 1408.6836}}].

\bibitem{Banfi:2013eda}
A.~Banfi, P.~F. Monni and G.~Zanderighi, \emph{{Quark masses in Higgs
  production with a jet veto}},
  \href{https://doi.org/10.1007/JHEP01(2014)097}{\emph{JHEP} {\bfseries 01}
  (2014) 097}, [\href{https://arxiv.org/abs/1308.4634}{{\ttfamily 1308.4634}}].

\bibitem{Hamilton:2015nsa}
K.~Hamilton, P.~Nason and G.~Zanderighi, \emph{{Finite quark-mass effects in
  the NNLOPS POWHEG+MiNLO Higgs generator}},
  \href{https://doi.org/10.1007/JHEP05(2015)140}{\emph{JHEP} {\bfseries 05}
  (2015) 140}, [\href{https://arxiv.org/abs/1501.04637}{{\ttfamily
  1501.04637}}].

\bibitem{Frederix:2016cnl}
R.~Frederix, S.~Frixione, E.~Vryonidou and M.~Wiesemann, \emph{{Heavy-quark
  mass effects in Higgs plus jets production}},
  \href{https://doi.org/10.1007/JHEP08(2016)006}{\emph{JHEP} {\bfseries 08}
  (2016) 006}, [\href{https://arxiv.org/abs/1604.03017}{{\ttfamily
  1604.03017}}].

\bibitem{Neumann:2016dny}
T.~Neumann and C.~Williams, \emph{{The Higgs boson at high $p_T$}},
  \href{https://doi.org/10.1103/PhysRevD.95.014004}{\emph{Phys. Rev.}
  {\bfseries D95} (2017) 014004},
  [\href{https://arxiv.org/abs/1609.00367}{{\ttfamily 1609.00367}}].

\bibitem{Boughezal:2013uia}
R.~Boughezal, F.~Caola, K.~Melnikov, F.~Petriello and M.~Schulze, \emph{{Higgs
  boson production in association with a jet at next-to-next-to-leading order
  in perturbative QCD}},
  \href{https://doi.org/10.1007/JHEP06(2013)072}{\emph{JHEP} {\bfseries 06}
  (2013) 072}, [\href{https://arxiv.org/abs/1302.6216}{{\ttfamily 1302.6216}}].

\bibitem{Chen:2014gva}
X.~Chen, T.~Gehrmann, E.~W.~N. Glover and M.~Jaquier, \emph{{Precise QCD
  predictions for the production of Higgs + jet final states}},
  \href{https://doi.org/10.1016/j.physletb.2014.11.021}{\emph{Phys. Lett.}
  {\bfseries B740} (2015) 147--150},
  [\href{https://arxiv.org/abs/1408.5325}{{\ttfamily 1408.5325}}].

\bibitem{Boughezal:2015aha}
R.~Boughezal, C.~Focke, W.~Giele, X.~Liu and F.~Petriello, \emph{{Higgs boson
  production in association with a jet at NNLO using jettiness subtraction}},
  \href{https://doi.org/10.1016/j.physletb.2015.06.055}{\emph{Phys. Lett.}
  {\bfseries B748} (2015) 5--8},
  [\href{https://arxiv.org/abs/1505.03893}{{\ttfamily 1505.03893}}].

\bibitem{Boughezal:2015dra}
R.~Boughezal, F.~Caola, K.~Melnikov, F.~Petriello and M.~Schulze, \emph{{Higgs
  boson production in association with a jet at next-to-next-to-leading
  order}}, \href{https://doi.org/10.1103/PhysRevLett.115.082003}{\emph{Phys.
  Rev. Lett.} {\bfseries 115} (2015) 082003},
  [\href{https://arxiv.org/abs/1504.07922}{{\ttfamily 1504.07922}}].

\bibitem{Chen:2016zka}
X.~Chen, J.~Cruz-Martinez, T.~Gehrmann, E.~W.~N. Glover and M.~Jaquier,
  \emph{{NNLO QCD corrections to Higgs boson production at large transverse
  momentum}}, \href{https://doi.org/10.1007/JHEP10(2016)066}{\emph{JHEP}
  {\bfseries 10} (2016) 066},
  [\href{https://arxiv.org/abs/1607.08817}{{\ttfamily 1607.08817}}].

\bibitem{Caola:2016upw}
F.~Caola, S.~Forte, S.~Marzani, C.~Muselli and G.~Vita, \emph{{The Higgs
  transverse momentum spectrum with finite quark masses beyond leading order}},
  \href{https://doi.org/10.1007/JHEP08(2016)150}{\emph{JHEP} {\bfseries 08}
  (2016) 150}, [\href{https://arxiv.org/abs/1606.04100}{{\ttfamily
  1606.04100}}].

\bibitem{Braaten:2017lxx}
E.~Braaten, H.~Zhang and J.-W. Zhang, \emph{{Mass Dependence of Higgs
  Production at Large Transverse Momentum}},
  \href{https://doi.org/10.1007/JHEP11(2017)127}{\emph{JHEP} {\bfseries 11}
  (2017) 127}, [\href{https://arxiv.org/abs/1704.06620}{{\ttfamily
  1704.06620}}].

\bibitem{Campbell:1999ah}
J.~M. Campbell and R.~K. Ellis, \emph{{An Update on vector boson pair
  production at hadron colliders}},
  \href{https://doi.org/10.1103/PhysRevD.60.113006}{\emph{Phys. Rev.}
  {\bfseries D60} (1999) 113006},
  [\href{https://arxiv.org/abs/hep-ph/9905386}{{\ttfamily hep-ph/9905386}}].

\bibitem{Campbell:2011bn}
J.~M. Campbell, R.~K. Ellis and C.~Williams, \emph{{Vector boson pair
  production at the LHC}},
  \href{https://doi.org/10.1007/JHEP07(2011)018}{\emph{JHEP} {\bfseries 07}
  (2011) 018}, [\href{https://arxiv.org/abs/1105.0020}{{\ttfamily 1105.0020}}].

\bibitem{Campbell:2015qma}
J.~M. Campbell, R.~K. Ellis and W.~T. Giele, \emph{{A Multi-Threaded Version of
  MCFM}}, \href{https://doi.org/10.1140/epjc/s10052-015-3461-2}{\emph{Eur.
  Phys. J.} {\bfseries C75} (2015) 246},
  [\href{https://arxiv.org/abs/1503.06182}{{\ttfamily 1503.06182}}].

\bibitem{Kudashkin:2017skd}
K.~Kudashkin, K.~Melnikov and C.~Wever, \emph{{Two-loop amplitudes for
  processes $g g \to H g, q g \to H q$ and $q \bar{q} \to H g$ at large Higgs
  transverse momentum}},  \href{https://arxiv.org/abs/1712.06549}{{\ttfamily
  1712.06549}}.

\bibitem{Catani:1998bh}
S.~Catani, \emph{{The Singular behavior of QCD amplitudes at two loop order}},
  \href{https://doi.org/10.1016/S0370-2693(98)00332-3}{\emph{Phys. Lett.}
  {\bfseries B427} (1998) 161--171},
  [\href{https://arxiv.org/abs/hep-ph/9802439}{{\ttfamily hep-ph/9802439}}].

\bibitem{Dulat:2015mca}
S.~Dulat, T.-J. Hou, J.~Gao, M.~Guzzi, J.~Huston, P.~Nadolsky et~al.,
  \emph{{New parton distribution functions from a global analysis of quantum
  chromodynamics}},
  \href{https://doi.org/10.1103/PhysRevD.93.033006}{\emph{Phys. Rev.}
  {\bfseries D93} (2016) 033006},
  [\href{https://arxiv.org/abs/1506.07443}{{\ttfamily 1506.07443}}].

\end{thebibliography}\endgroup

\end{document}